\documentclass[apj]{emulateapj}
\slugcomment{\sc{submitted to the Astrophysical Journal:} September 23, 2013}
\usepackage{graphicx}
\usepackage{natbib}
\usepackage[caption=false]{subfig}

\def\be{\begin{equation}}
\def\ee{\end{equation}}
\def\bea{\begin{eqnarray}}
\def\eea{\end{eqnarray}}

\begin{document}

\title{To Stack or Not to Stack: \\ 
Spectral Energy Distribution Properties of Ly$\alpha$-Emitting Galaxies at \lowercase{$z$} $=$ 2.1}

\author{Carlos J. Vargas\altaffilmark{1,2}, Hannah Bish\altaffilmark{1}, Viviana Acquaviva\altaffilmark{3}, Eric Gawiser\altaffilmark{1},
Steven L. Finkelstein\altaffilmark{4}, 
Robin Ciardullo\altaffilmark{5,6}, 
Matthew L. N. Ashby\altaffilmark{7}, 
John Feldmeier\altaffilmark{8}, 
Henry Ferguson\altaffilmark{9}, 
Caryl Gronwall\altaffilmark{5,6}, 
Lucia Guaita\altaffilmark{10}, 
Alex Hagen\altaffilmark{5,6}, 
Anton Koekemoer\altaffilmark{9}, 
Peter Kurczynski\altaffilmark{1}, 
Jeffrey A. Newman\altaffilmark{11},
\& Nelson Padilla\altaffilmark{12}}

\altaffiltext{1}{Department of Physics and Astronomy, Rutgers, The State University of New Jersey, Piscataway, NJ 08854}
    \altaffiltext{2}{Department of Astronomy, New Mexico State University, Las Cruces, NM 88001}     
        \altaffiltext{3}{Physics Department, New York City College of Technology, City University of New York, 300 Jay Street, Brooklyn, NY 11201}
              \altaffiltext{4}{Department of Astronomy, The University of Texas at Austin, Austin, TX 78712}
              \altaffiltext{5}{Department of Astronomy \& Astrophysics, The Pennsylvania State University, University Park, PA 16802}
              \altaffiltext{6}{Institute for Gravitation and the Cosmos, The Pennsylvania State University, University Park, PA 16802}
              \altaffiltext{7}{Harvard-Smithsonian Center for Astrophysics, 60 Garden St., Cambridge, MA 02138} 
              \altaffiltext{8}{Department of Physics and Astronomy, Youngstown State University, Youngstown, OH 44555}
          \altaffiltext{9}{Space Telescope Science Institute, Baltimore, MD}
\altaffiltext{10}{Department of Astronomy, Oskar Klein Center, Stockholm University, Roslagstullsbacken 21, Stockholm, Sweden}
              \altaffiltext{11}{Department of Physics and Astronomy, University of Pittsburgh, 3409 O'Hara St., Pittsburgh, PA 15260}  
\altaffiltext{12}{Departmento de Astronomia y Astrofisica, Universidad Cat\'{o}lica de Chile, Santiago, Chile}

\begin{abstract}

We use the Cosmic Assembly Near-Infrared Deep Extragalactic Legacy
Survey (CANDELS) GOODS-S multi-wavelength catalog to identify
counterparts for 20 Ly$\alpha$ Emitting (LAE) galaxies at $z =
2.1$. We build several types of stacked Spectral Energy Distributions
(SEDs) of these objects. We combine photometry to form average and
median flux-stacked SEDs, and postage stamp images to form average and
median image-stacked SEDs. We also introduce scaled flux stacks that
eliminate the influence of variation in overall brightness. We use the
SED fitting code SpeedyMC to constrain the physical properties of
individual objects and stacks. Our LAEs at $z = 2.1$ have stellar
masses ranging from $2\times 10^{7}$ $M_{\odot}$ - $8\times 10^{9}$
$M_{\odot}$ (median $= 3\times 10^8 M_{\odot}$), ages ranging from 4
Myr to 500 Myr (median =100 Myr), and E(B-V) between 0.02 and 0.24
(median = 0.12). We do not observe strong correlations between
Ly$\alpha$ equivalent width (EW) and stellar mass, age, or E(B-V). The
Ly$\alpha$ radiative transfer ($q$) factors of our sample are
predominantly close to one and do not correlate strongly with EW or
E(B-V), implying that Ly$\alpha$ radiative transfer prevents
Ly$\alpha$ photons from resonantly scattering in dusty regions. The
SED parameters of the flux stacks match the average and median values
of the individual objects, with the flux-scaled median SED performing
best with reduced uncertainties.  Median image-stacked SEDs provide a
poor representation of the median individual object, and none of the
stacking methods captures the large dispersion of LAE properties. 

\end{abstract}


\newpage
\section{Introduction}

Lyman Alpha Emitting (LAE) galaxies at $2<z<3$ are thought to be
building blocks of present-day galaxies like our own Milky Way
\citep{gawiseretal07,guaitaetal11}. These galaxies are easy to detect
thanks to the strength of the Lyman-$\alpha$ (Ly$\alpha$) line, but
they usually have low stellar masses and are therefore dim in the
continuum. As a result, it has been a common procedure to stack
multiple photometric data points of LAEs in order to enhance their
signal-to-noise ratio
\citep{gawiseretal06b,gawiseretal07,nilssonetal07,pirzkaletal07,finkelsteinetal07,laietal08,pentericcietal08,onoetal10a,onoetal10b,ouchietal10,nilssonetal11,guaitaetal11,finkelsteinetal11b,acquavivaetal11,acquavivaetal12a}.

The nature of radiation emitted from astrophysical sources offers
insights into their physical properties. Ly$\alpha$ photons are
emitted when a neutral hydrogen atom undergoes a transition from the
first excited state to the ground state. An immense number of
Ly$\alpha$ photons are emitted at 1216 $\AA$ rest frame from HII
regions surrounding young, massive stars. The large luminosity of the
Ly$\alpha$ emission line makes it easily identified even over vast
distances and thus it is widely used in research in extragalactic
astronomy. LAEs are typically discovered using the narrow-band filter
technique \citep{cowieh98,rhoadsetal00}, in which a detection of an
LAE is made by finding greater signal to noise ratios in a narrow-band
filter image than in images obtained with adjacent broad-band
filters. 

Inconsistencies in the literature, particularly in estimating ages and
sample diversity, motivate a study of the validity of the stacking
method for analyzing the Spectral Energy Distributions (SEDs) of dim
LAEs. To date, most SED studies of high-redshift LAEs have been done
by stacking the catalog fluxes or ground-based postage stamp images of
galaxies too faint to study individually. By stacking numerous faint
objects (via average or median) we increase the signal to noise
significantly. It is believed that a study of these images will
represent the characteristics of a sample of galaxies on the average.
However, this interpretation of median as typical relies on the
assumption that the best fit SED parameters of the stacked galaxies
have similar physical properties. An additional key assumption is that
the median (average) SEDs will match the median (average) physical
properties of individual galaxy SEDs. This assumption appears
reasonable for stellar mass if the mass to light ratio of the stellar
populations of different LAEs is similar, but it is not guaranteed for
other parameters like age or dust. In conjunction with this study, a
study by \citet{hagenetal13}, in which LAEs identified by the
Hobby-Eberly Telescope Dark Energy eXperiment (HETDEX) pilot survey
are studied individually through SED fitting, could shed more light on
these inconsistencies.

Results from the analysis of stacked samples of LAEs indicate a large
spread in the properties of these galaxies and introduce further
concern about the validity of the stacking technique.  
A recent study of
median image-stacked LAEs at z = 3.1 found them to be older than
median flux-stacked LAEs at z = 2.1 \citep{acquavivaetal12b}; this
puzzling result would imply that LAEs grow younger as cosmic time
progresses. However, it might instead result from the failure of image
and/or flux stacking to accurately measure the SED characteristics of
individual galaxies, and from systematic differences between
image-stacking and flux-stacking. Additionally, a previous study of
individual LAEs at $z=4.5$ \citep{finkelsteinetal09} found a wide
range of physical properties, including many LAEs which appeared
dusty, something that would not have been expected given the earlier
results at $z=3.1$ from \citet{gawiseretal06a} which implied, via a stacking
analysis, that LAEs were dust free.

In this paper, we perform SED fitting on 20 LAEs at $z = 2.1$ as well as
on their flux- and image-stacked SEDs, in order to clarify whether
these discrepancies are attributable to the failure of one or all
stacking methods. The use of deep HST and Spitzer IRAC data gives us
the opportunity, for the first time, to study rest-frame ultra-violet
and optical properties of individual objects, which is crucial for
accurate constraints on age and stellar mass. We use the Markov Chain
Monte Carlo (MCMC) algorithm ¡ÈSpeedyMC¡É \citep{acquavivaetal12b} to
analyze the SED properties of our sample. SpeedyMC is a streamlined
implementation of GalMC \citep{acquavivaetal11} designed to handle
large samples of SEDs. This code utilizes Bayesian statistics to
determine the expectation values of SED parameters and provide
estimates of the uncertainties associate to their measurements. All
SED fitting calculations assume a WMAP-based cosmology including $H_0
= 73$~km~s$^{-1}$~Mpc$^{-1}$

\section{Data}

The starting point for our analysis was the sample of 216 LAE
candidates at $z=2.1$ discovered in deep 3727 {\AA} narrow-band images
of the 30$^{\prime}$ $\times$ 30$^{\prime}$ Extended Chandra Deep
Field-South (ECDF-S) by \citet{guaitaetal10} as refined by
\citet{bondetal12}, who excluded 34 additional LAE candidates that
appear to be low-redshift contaminants due to extended morphology in
\textit{HST} images. These objects typically have low signal to noise
ratios in ground-based broad-band images, with a median magnitude of
$R\sim 25.5$ and are unresolved at 0.1 arcseconds. Thus, the addition
of deep \textit{HST} imaging is crucial.

CANDELS has produced a deep H-band-selected multiwavelength catalog of
the GOODS-S field, which encompasses the central $16^\prime \times
10^\prime$ of ECDF-S \citep{guoetal13}. The observations are 10 epochs
deep in H band, which corresponds to roughly six \textit{HST} orbits.
The catalog contains photometry from $U$-band images from CTIO and
VIMOS, ACS ($BViz$) images from the GOODS survey
\citep{giavaliscoetal04}, CANDELS WFC3 images (F098M,$YJH$), $K$-band
images from Hawk-I and ISAAC, and {\it Spitzer Space Telescope}
Infrared Array Camera (IRAC; \citealt{fazioetal04}) 3.6, 4.5 $\mu$m
images from SEDS \citep{ashbyetal13} and 5.8, 8.0 $\mu$m images from GOODS
\citep{dickinsonetal03b}.  For SED fitting, we excluded both U bands
due to expected Ly$\alpha$ contamination at this redshift, F098M
due to having coverage in a minority of our sample, and both K-bands
due to incomplete coverage and shallower photometry.  All of our
analysis made use of imaging and photometry from the remaining 11
space-based bands.

We match the positions of our catalog of known LAEs at z=2.1 to the
positions of objects contained within the version of the CANDELS
Multi-wavelength catalog presented in \citet{guoetal13}. Objects
within a distance of $0.5$ arcseconds were recorded as a match, and
their CANDELS SEDs were used in our analysis. Only two objects showed
multiple possible counterparts within this tolerance, and they were
discarded due to the strong likelihood of neighboring object
contamination.  In total, we found 20 LAE counterparts in the CANDELS
Multi-wavelength catalog. We note that only 30 of the original 216
LAEs are within the GOODS-S field, and the 10 ``missing'' objects were
probably too dim to have been detected at the 5$\sigma$ level in H
band, and were thus excluded from the CANDELS catalog.

\begin{figure*}[ht!]
\centering
\includegraphics[scale=0.8]{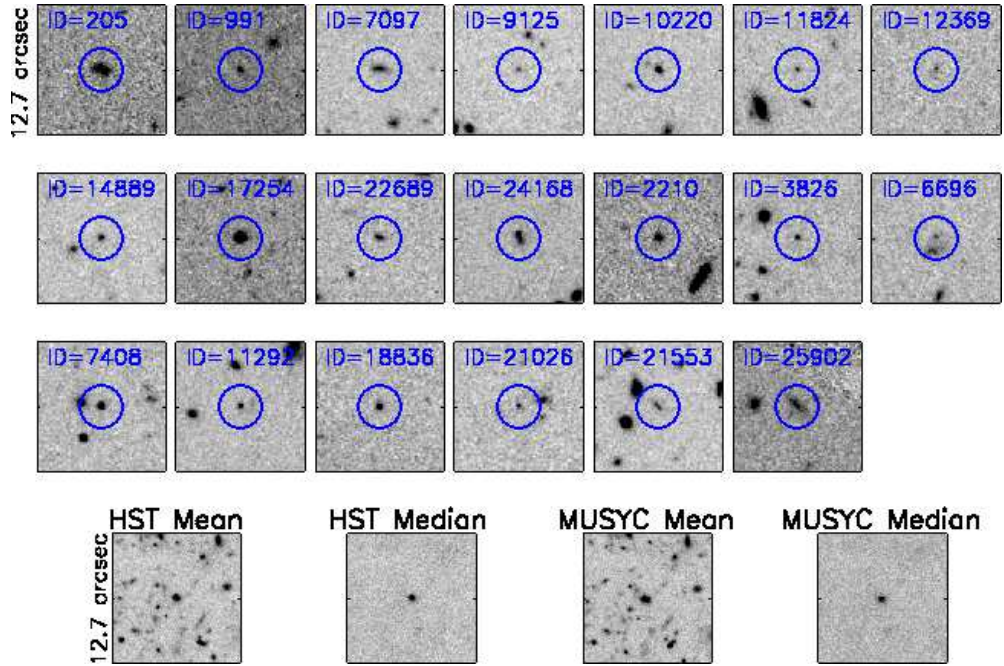}
\caption{HST F160W filter postage stamp images of the 20 individual
  objects and image stacks in our sample. \textit{HST} stacks are
  obtained from centroiding on the CANDELS object positions, while
  \textit{NB}-centered stacks are obtained from centroiding on the
  MUSYC \textit{NB} positions. The blue circles highlight the objects
  to aid the eye. In these stamps, North points to the top of the
  page, and East points to the right. All Stamp images are square of
  side length $12.7\arcsec$.\label{fig1} }
\end{figure*}

The main objective of this paper is the comparison of the SED fitting
properties of individual objects and stacks. The SEDs of the stacks
can be obtained via two different procedures: by combining the
cataloged fluxes of the individual objects, or by stacking the
postage-stamp images of the 20 individual objects at each wavelength
and then performing photometry on the combined images. We refer to the
SEDs obtained via these two methods as ``flux-stacked'' and
``image-stacked'' respectively.  We took the median and average of the
flux densities in each band of all 20 $z=2.1$ LAEs with CANDELS
counterparts and will henceforth refer to these as the median and
average flux stacks respectively.

To create the image-stacked SED, we performed both average and median
stacking from cutout images centered around each of our 20 LAEs.  The
extracted \textit{HST} cutouts measured 211 pixels on a side, while
the IRAC cutouts measured 21 pixels.  As the processed \textit{HST}
image pixels measure 0.06$^{\prime \prime}$ and the processed IRAC
image pixels measure 0.6$^{\prime \prime}$, the cutouts in all bands
had approximately the same size of 12.6$^{\prime \prime}$ on a side.
We note that of the 20 objects, 6 of them were located in the northern
$\sim$25\% of the GOODS-S field, which has WFC3 data from the WFC3
Early Release Science program (ERS; \citealt{windhorstetal11}).  While
these images were obtained prior to the CANDELS data, they were
reprocessed alongside the CANDELS data to make a single mosaic which
was used to extract the cutouts.  Along with the science images, we
also extracted cutouts from the r.m.s. images for use with Source
Extractor (\citealt{bertina96}, see \citealt{koekemoeretal11} for details
on the science and r.m.s. mosaic construction).

To create our image stacks, we created median and average images by
taking the median/average value for each pixel from all of the cutout
images for a given band, excluding pixels with r.m.s. values $>$ 10; a
very high value, indicative of a problem with the given pixel in that
image.  We then measured photometry on the image stacks using Source
Extractor in dual-image mode, with the F160W stack as the detection
image for the \textit{HST} bands and the 3.6 $\mu$m image as the
detection band for the IRAC images.  We measured photometry using
Kron-style elliptical apertures, denoted by MAG\_AUTO within Source
Extractor.  For the \textit{HST} bands, we measured colors in small
Kron apertures (with the parameter PHOT\_AUTOPARAMS set to 1.2, 1.7),
with the fluxes corrected to total using an aperture correction
derived by the ratio of the default MAG\_AUTO flux value to our
smaller Kron flux (this procedure has been optimized for faint,
high-redshift sources; see \citealt{finkelsteinetal10} for details).
For the IRAC stacks, we simply measured photometry in the larger Kron
aperture, which is designed to approximate a total flux. The
photometry for the individual objects was also performed in the same
way.

\begin{figure}[ht!]
\centering
\includegraphics[scale=0.47]{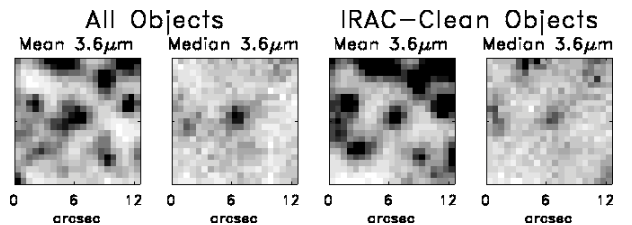}
\caption{Average and median postage stamp images for the entire sample
and the sample deemed clean in the IRAC $3.6 \mu m$ band. Note the
reduced signal to noise in the clean median image; an effect of its
lower sample size.\label{fig2}}
\end{figure}

To further our analysis we also introduce an additional set of image
stacks centered on the object positions found in the MUSYC narrow-band
(\textit{NB}) catalog of LAEs. Comparison of the primary HST-centered
image stacks with these \textit{NB}-centered stacked images will
illustrate any improvement that results from eliminating the
$\sim0.1''$ narrow-band centroiding errors \citep{guaitaetal10} and
any astrometric offsets between ground-based and space-based
coordinates. Figure \ref{fig1} shows F106W filter postage stamp images
of each galaxy in our sample and those of the \textit{HST} and
\textit{NB}-centered image stacks. In order to explore the effects of
potential IRAC contamination, we introduce IRAC-clean image
stacks. These stacks were comprised of ten individual objects visually
deemed clean in the IRAC 3.6$\mu m$ band. The IRAC-clean stack postage
stamp images are compared to those of the entire sample in Figure
\ref{fig2}. A slight loss of S/N in the clean sample is apparent due
to the reduced sample size.  Figure \ref{fig3} shows the SEDs of all
individual objects and all stacks overlaid atop one-another for
comparison. We also introduce a subsample of our 20 LAEs that were
visually deemed ``clean'', or free of nearby sources in the IRAC $3.6
\mu m$ band. Of the original 20 LAEs in our sample, 11 of them
comprise the IRAC-clean sample.

\subsection{Error estimation}
The uncertainty in the fluxes for individual objects' SEDs
include both a photometric error and a zero-point error arising from
the combination of calibration and zero-point errors; these are added
in quadrature. We assume the calibration error to be 3\% in
\textit{HST} bands and 8\% for the IRAC bands (CANDELS team, private
communication).

Errors on the flux-stacked photometry are determined by a bootstrap
procedure, as in \citet{guaitaetal11} and
\citet{acquavivaetal11}. This accounts for both the photometric error
and the sample variance resulting from the spread in the SEDs of
different galaxies; for our sample, the latter is the dominant source
of uncertainty. The same calibration uncertainty used for the
individual objects' SED is added in quadrature to the bootstrap error
in both the average and median flux stacked SED. For image-stacked
SEDs, we adopt a conservative error estimate in each band by taking
the larger of the photometric error indicated by SExtractor and the
bootstrap error, and adding it in quadrature to the calibration error.

\subsection{Stacking Simulations}

In order to determine the expected impact of flux- and image-stacking
on photometry in a given band, we conducted a series of Monte Carlo
simulations.  We modeled the behavior in WFC3 NIR images, with 0.06$"$
pixels sampling a 0.18$"$ PSF.  A realistic power-law flux
distribution was assigned, with fluxes ranging over a factor of 30
distributed uniformly in log(flux).  The median of this distribution
is only 63\% of its average.  We then varied the relative level of
errors in centroiding, background subtraction, and photometry to check
their influence on the resulting average- and median-stacked
fluxes. The typical circular aperture used for photometry recovers
71\% of the input flux, and we assume hereafter that this can be
corrected to produce an unbiased estimate of total flux.  Doing so,
without any sources of error,
both flux stacks and \textit{HST}-centered image stacks are unbiased
in their estimations of median and average flux.
Adding a low level of photometric error so that the median
object is detected at $S/N \sim 10$ makes no significant difference;
this is realistic for H-band where the catalog was detected.

\begin{figure*}
\centering
\includegraphics[scale=1.0]{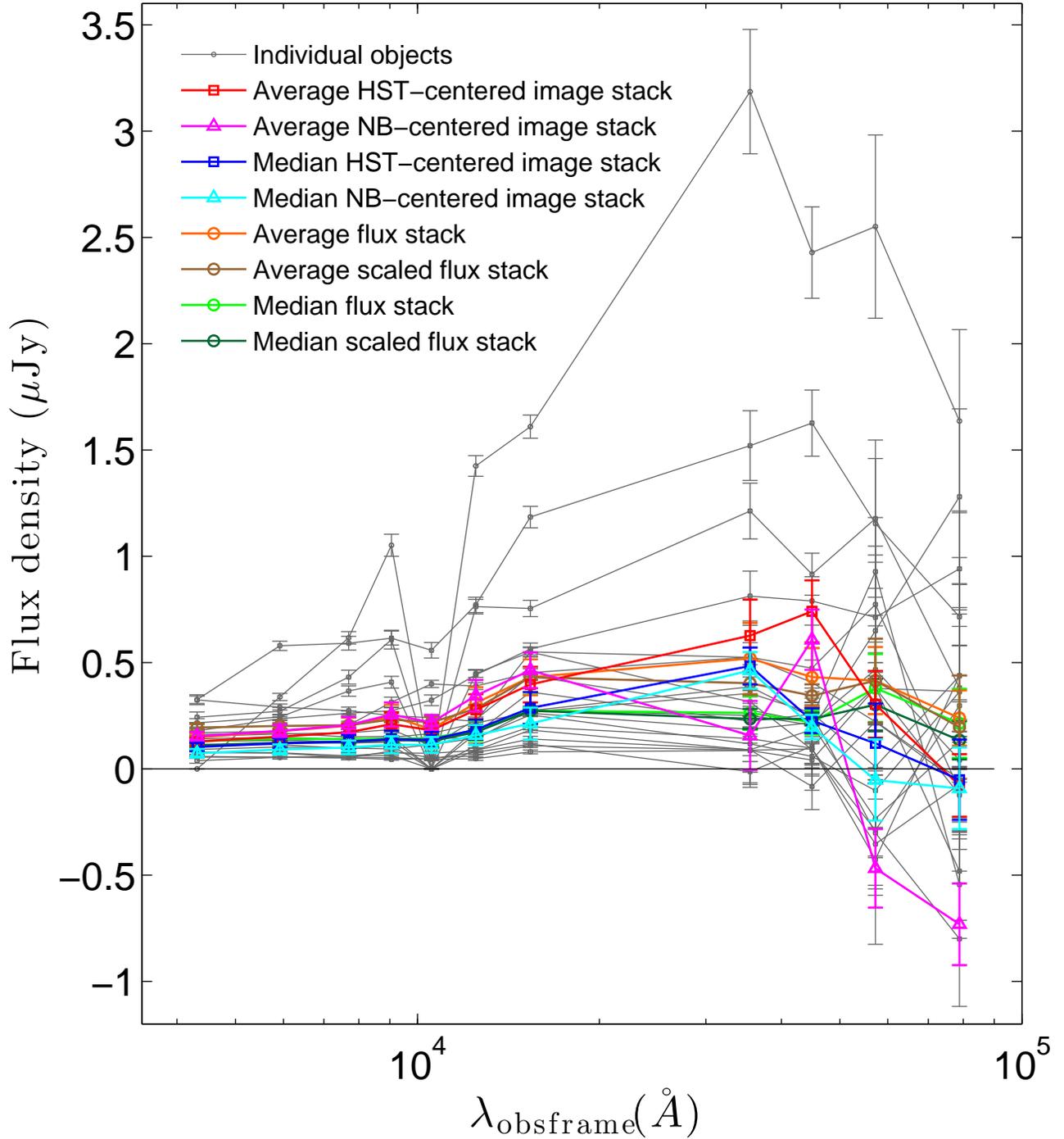}
\caption{Above, the SEDs of each individual object (gray) are plotted
  with all of the stacked SEDs in our analysis. Red is the average
  \textit{HST} image stack, blue is the median \textit{HST} image
  stack, magenta is the average NB-centered image stack, cyan is the
  median NB-centered image stack, green is the median flux stack,
  orange is the average flux stack, brown is the average scaled flux
  stack, and dark green is the median scaled flux stack. One can
  observe the median flux stack's robustness to outlying sources,
  especially in the \textit{Spitzer IRAC} bands. \label{fig3}}
\end{figure*}

However, centroiding errors are inevitable; for CANDELS objects
detected at $S/N > 5$, these errors should be at most a single pixel
\citep{koekemoeretal12}.  When we simulate 1 pixel r.m.s. centroiding
errors but no additional sources of error, the average flux and
average HST-centered image stacks lose 24\% of the flux, the median
flux stack loses 29\% and the median HST-centered image stack loses
43\% of the flux.  Adding realistic levels of photometric and
background subtraction errors makes no significant difference in these
results.  The larger flux loss in median image stacks may explain why
these are the dimmest stacked SEDs at wavelengths imaged by HST, as
seen in Figure \ref{fig3}.

The overall stacking behavior is similar at the much larger pixel
scale of IRAC bands (e.g., 1.2$"$ pixels sampling a 1.8$"$ PSF at 3.6
microns), where we continued to assume that the centroiding errors
were no larger than 0.06$"$ given the usage of CANDELS H-band catalog
positions for the HST-centered image stacks and for TFIT photometry on
IRAC images used to make the multiwavelength catalog inputs for
flux-stacking.  These centroiding errors cause no measurable loss in
flux at IRAC resolution.  However, this could create a mild color bias
in the stacks since, as noted above,
24-43\% of the flux is being lost to centroiding errors in
\textit{HST} bands.  The worst effect is expected in the median
HST-centered image stack, which does exhibit a red $H-$[3.6] color in
Fig. 3.  It should be noted that our simulations do not include the
non-Gaussian PSF wings typical in IRAC bands or any attempt to account
for the contamination by neighboring objects that results from them.

We also simulated the NB-centered image stacks where the ground-based
MUSYC narrow-band-detected LAE positions were adopted, causing 0.1$"$
centroid errors.  If we had used $\sim$FWHM diameter apertures, which
are formally optimal for point sources, in the \textit{HST} images,
the NB-centered average (median) stacks would underestimate
\textit{HST}-band fluxes by a factor of 2 (6).  However, the larger
apertures used in $H$-band to determine aperture corrections are big
enough to correct this bias for the average NB-centered image stack,
which is well worth the loss of S/N.

It is important to note that no matter what aperture size is used, our
simulations show that {\em median image stack} photometry is biased
low by a large factor whenever centroiding errors are significant
compared to the PSF.  This occurs because the various objects exhibit
only partial overlap, leading median image values to be dominated by
image background rather than objects.  This provides a significant
note of caution about the usage of median image stacks, although the
effect appears to be modest for our NB-centered median image stack and
should be negligible for our HST-centered median image stack due to
the small expected centroiding errors.

\section{SED Fitting Methodology}

Information about the physical properties of galaxies, including
redshift, stellar mass, age, dust content, and metallicity, can be
determined by fitting their SEDs. The algorithm that was used for this
analysis, SpeedyMC, is a faster version of GalMC introduced in
\citet{acquavivaetal12b}. Rather than using GALAXEV to generate the
model SED at each point, a template library on a grid of locations
encompassing the entire parameter space is generated beforehand. The
exploration of parameter space is then carried out using the same MCMC
algorithm, but at each location, multi-linear interpolation between
the pre-computed spectra is used to calculate the model SED and its
corresponding $\chi^{2}$ value. The use of SpeedyMC method allows us
to fit the SEDs of galaxies at a rate 20,000 times faster than GalMC,
and corresponding to about one second per galaxy on a 2.2GHz MacBook
Pro laptop.

\begin{figure}[ht!]
\centering
\includegraphics[width=3.0in]{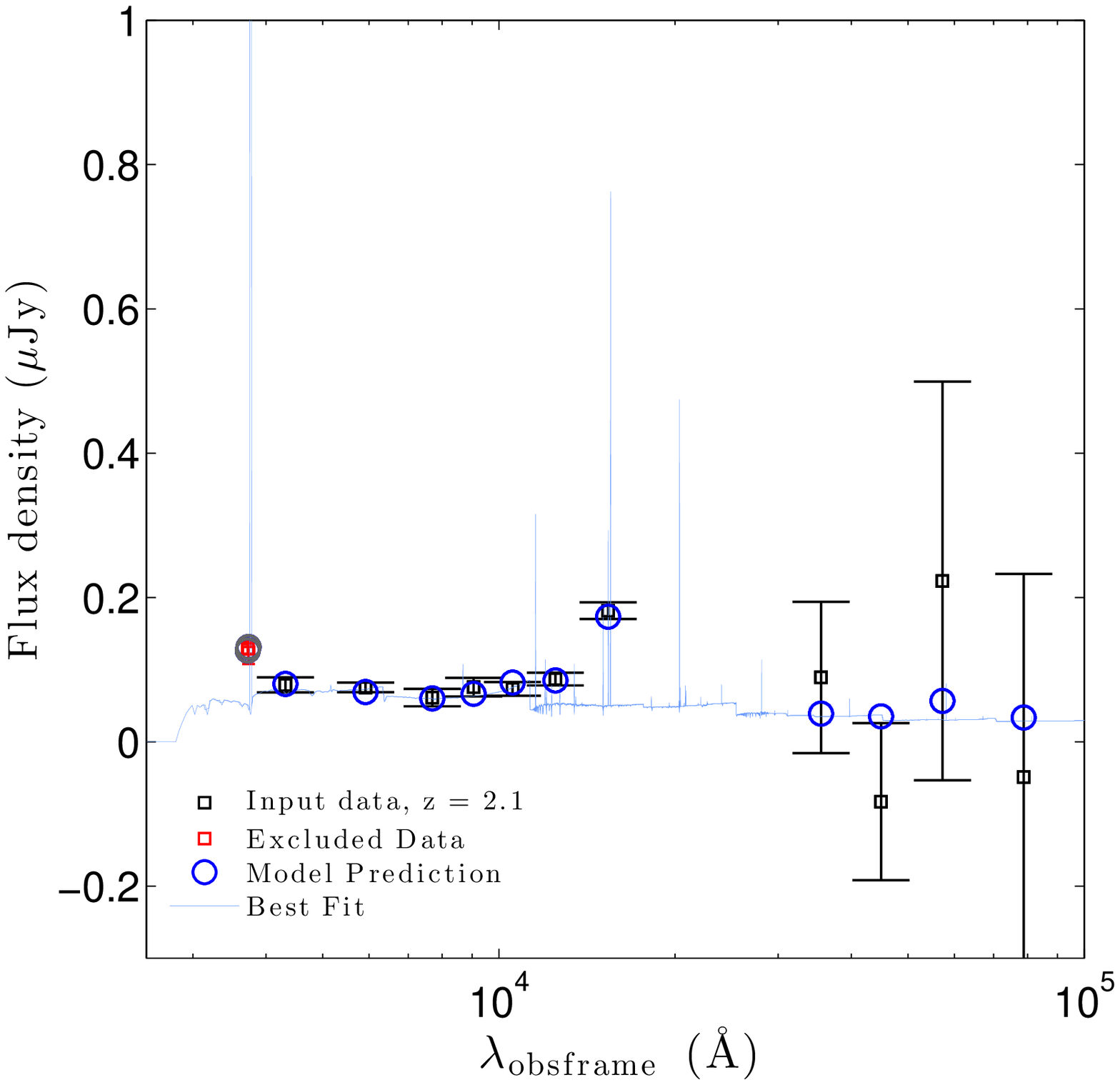}
\includegraphics[width=3.0in]{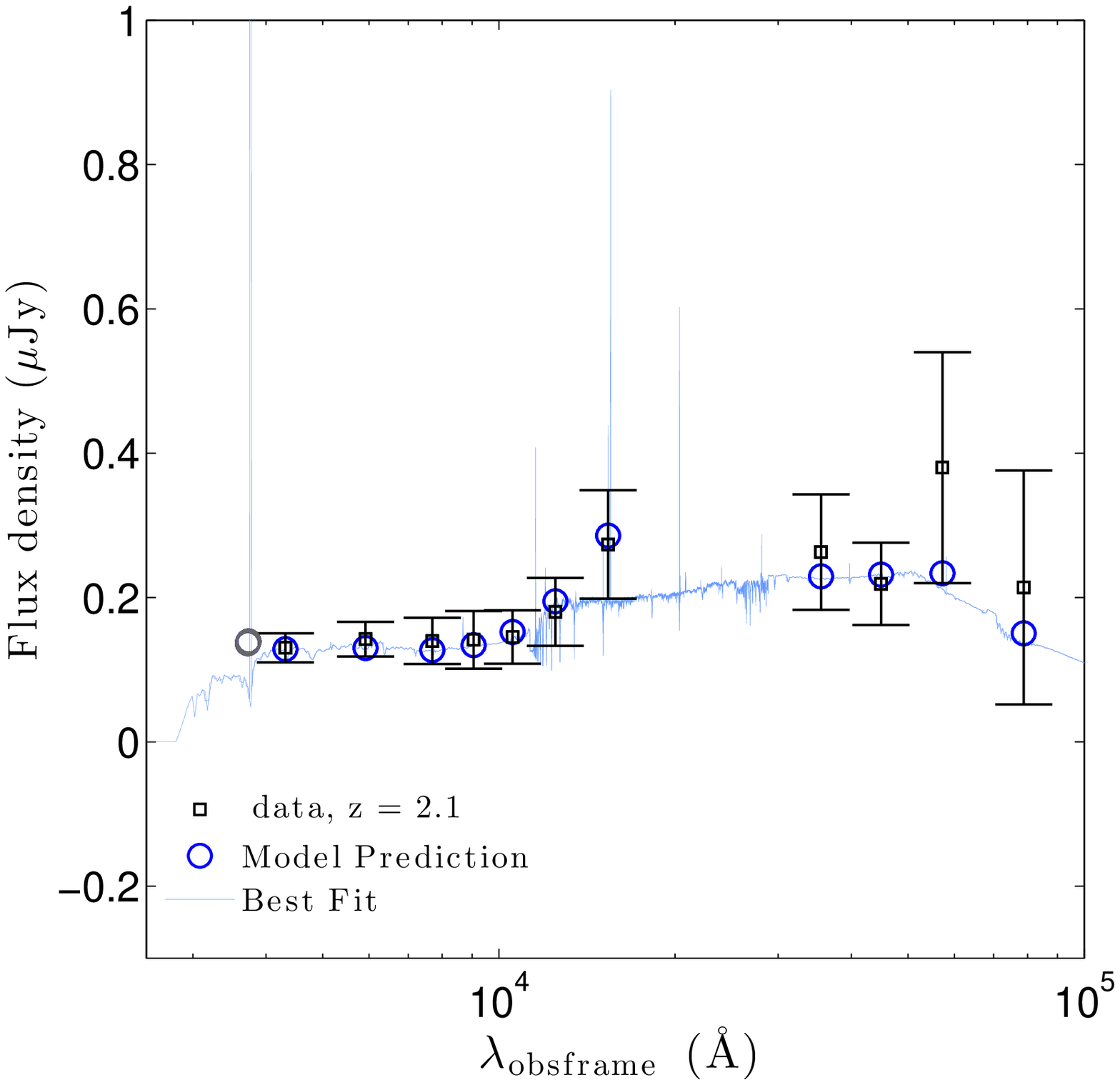}
\caption{{\it Top panel:}A plot of the SED of individual object 3826
  overlaid with its MCMC best fit SED template. These templates
  correspond to the model SED which produces the lowest $\chi ^{2}$
  value by SpeedyMC. Note the $U$-band data is affected by Ly$\alpha$
  emission and that the emission-line contribution is, in this case,
  predicted extremely well by the best-fit model to the 11 space-based
  bands used for SED fitting. {\it Bottom panel:} A similar plot for
  the median flux stack. \label{fig4}}
\end{figure}

In our analysis, the parameter space consisted of stellar mass, age,
and dust reddening defined by excess color E(B-V). We assumed a
constant star formation history, and fixed the metallicity, $Z$, at
the value of 0.2$Z_{\odot}$ as indicated by both spectroscopic
analysis and SED fitting results
\citep{acquavivaetal11,finkelsteinetal11}.  We used the stellar
population synthesis models of \citet{bruzualc03} (hereafter BC03) and
Charlot \& Bruzual (2007, Private Comm.)  (hereafter CB07), and
included nebular emission according to the procedure described in
\citet{acquavivaetal11}. Galactic absorption was taken into account
using the Calzetti law \citep{calzettietal94}, with a value $R_V =
4.05$, and starlight absorption by the IGM using the prescription from
\citet{madau95}. We used the Salpeter initial mass function for
consistency with the previous literature on this subject, and assumed
a WMAP cosmology. Our reference grid contained 100 values of both age
and E(B-V) (the stellar mass is a free normalization parameter which
can be varied without the need to build extra models in the grid). The
number of values in the grid was chosen by refining the resolution of
the grid until the resulting probability distributions obtained from
GalMC and SpeedyMC were virtually identical, a sign that the linear
interpolation between reference SED was working correctly. We limit
the size of the grid in age to be between $10^{6}$ and $3.2 \times
10^{9}$ years, as the universe at $z=2.1$ is $3.2 \times 10^{9}$ years
old. In mass, we limit the grid to range $10^{4}$ and $10^{15}$
$M_{\odot}$, and in E(B-V) we range $0$ to $1$.  We ran six chains on
each object, using $100,000$ steps for each and starting from six
different locations to ensure sampling of the entirety of parameter
space and minimize ``chain locking'' in local minima. The ``GetDist''
software from \citet{Lewis02} was used to analyze the chains and to
make sure that the distribution inferred from the chains had converged
to the true one. In the left panel of Figure \ref{fig4}, we plot the
observed SED of CANDELS object 3826 overlaid with its best-fitting
model produced by SpeedyMC. A similar plot for the median flux stacked
SED is included as the right panel of Figure \ref{fig4}. We plot the
observed U-band data points in this figure as an illustration of
possible Ly$\alpha$ contamination in that band.

\section{Results}

The BC03 stellar population synthesis models performed better than the
CB07 ones, resulting in smaller $\chi^2$ values on average. This
behavior is in agreement with results in the recent literature that
seemed to favor a low contribution of thermally pulsating asymptotic
giant branch stars (e.g., \citealt{krieketal10},
\citealt{meidtetal13}, \citealt{zibettietal13}). For these reasons, we
show the parameters obtained by using the BC03 models in Table 1 and
the figures. Individual LAEs at $z=2.1$ are found to have stellar
masses ranging from $2.3\times 10^{7}$ $M_{\odot}$ to $8.5\times
10^{9}$ $M_{\odot}$, ages ranging from $0.004$ Gyr to $0.47$ Gyr, and
E(B-V) between $0.02$ and $0.24$.

\begin{figure*}[ht!]
\centering
\subfloat[]{\includegraphics[width=2.2in]{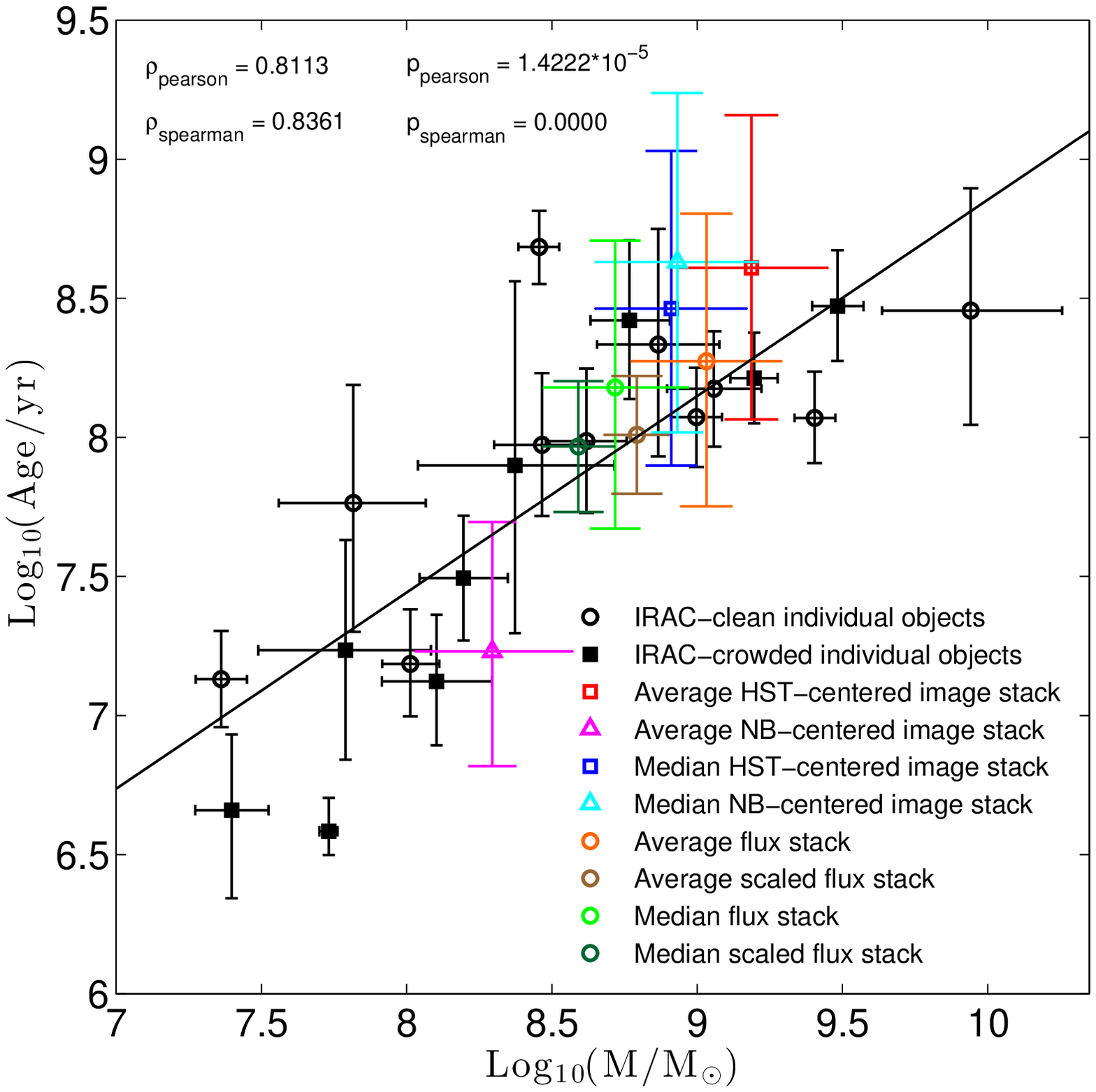}}
\subfloat[]{\includegraphics[width=2.2in]{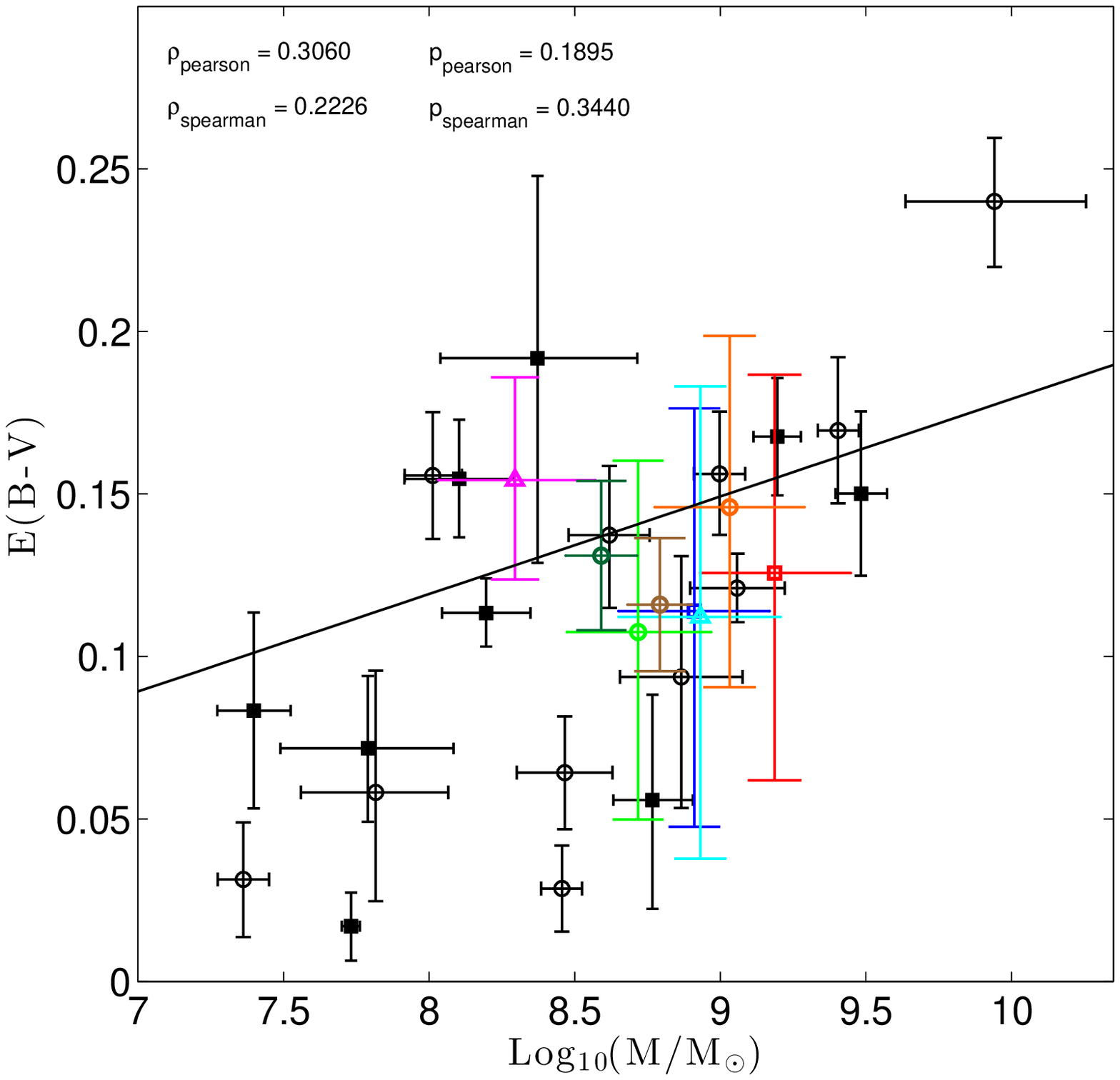}}
\subfloat[]{\includegraphics[width=2.2in]{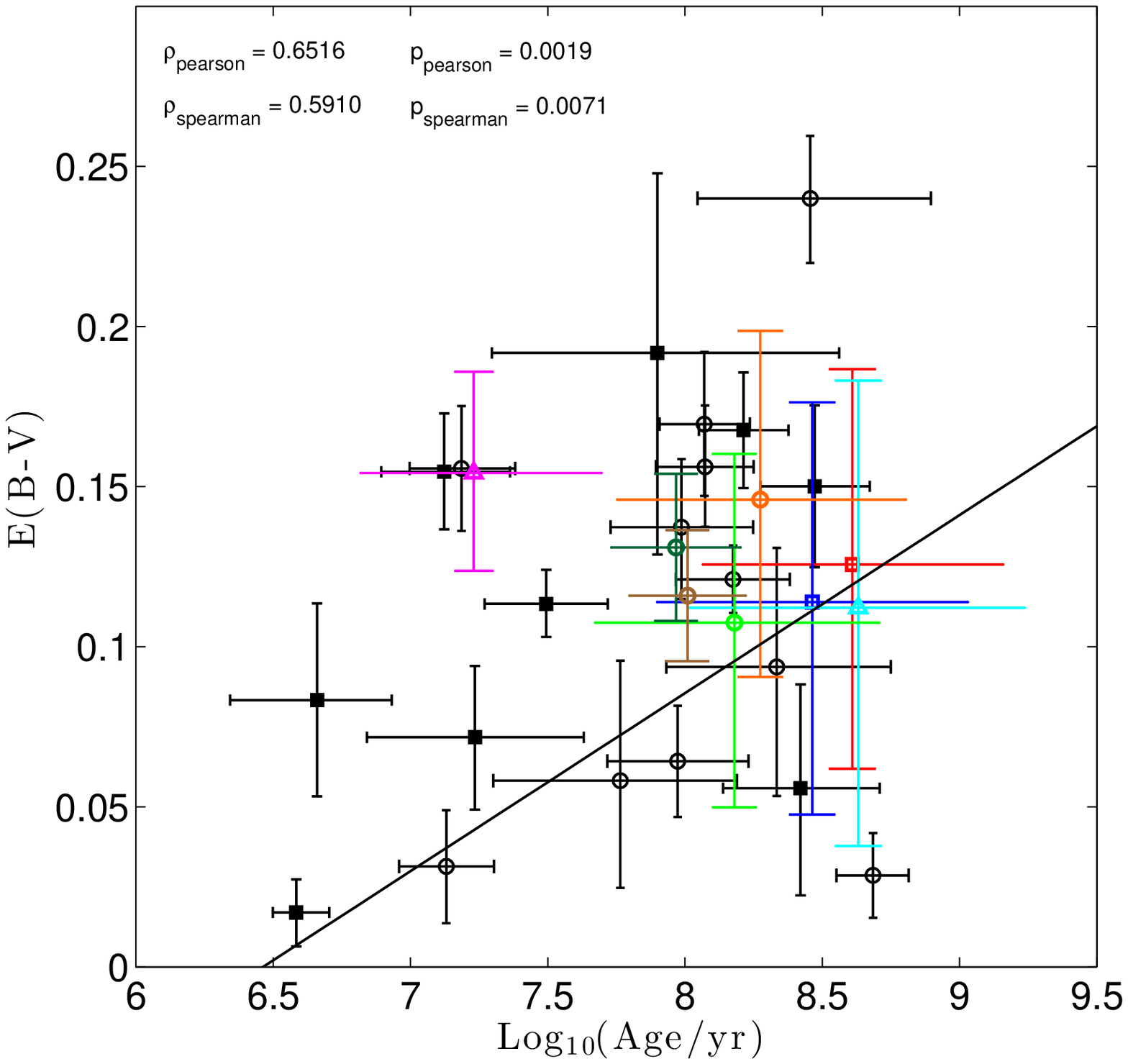}}
\caption{These plots show the relationships between (a) stellar mass
  and age, (b) stellar mass and E(B-V), and (c) age and E(B-V) of the
  sample. The black points represent the best fitting parameter values
  for the individual objects, and the remaining color scheme is
  retained from Figure \ref{fig3}. The error bars plotted are those output by
  SpeedyMC and are asymmetrical.\label{fig5}}
\end{figure*}

Figure \ref{fig5} shows the relationships between parameters of age,
stellar mass, and E(B-V). In our analyses of parameters which may have
statistically significant correlations with other parameters, we use
both the Pearson product-moment correlation coefficient and the
Spearman rank correlation coefficient.  The Pearson and Spearman
correlation coefficients ($\rho$) describe the strength and direction
of the correlation and range from $-1$ to $1$. We define $1-p$ as the
confidence with which one can reject the null hypothesis of no
correlation. In order for two parameters to be correlated at 95\%
confidence, their $p_{Spearman}$ value must be less than
$0.05$. Because the intrinsic distribution of parameters is not well
known, we adopt the Spearman $p$ values. We see a correlation of age
with stellar mass, as expected. The assumption of a constant star
formation history forces older galaxies to be more massive, and thus,
we see a correlation between age and stellar mass. No correlation is
found between E(B-V) and stellar mass. Also, there is an observed
correlation between age and E(B-V), implying that older LAEs generally
host more dust.

\begin{figure*}[ht!]
\centering
\subfloat[]{\includegraphics[width=2.2in]{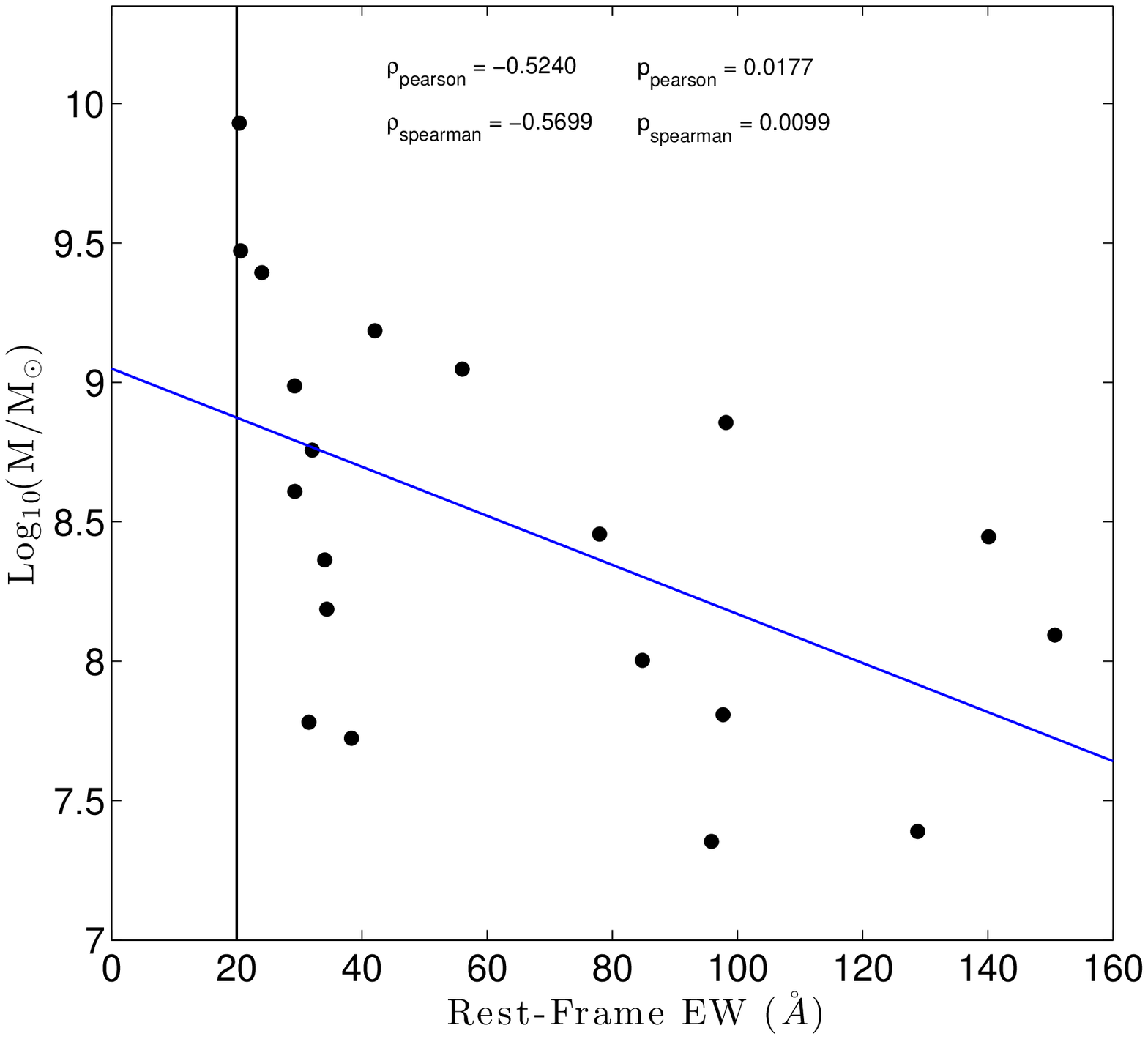}}
\subfloat[]{\includegraphics[width=2.2in]{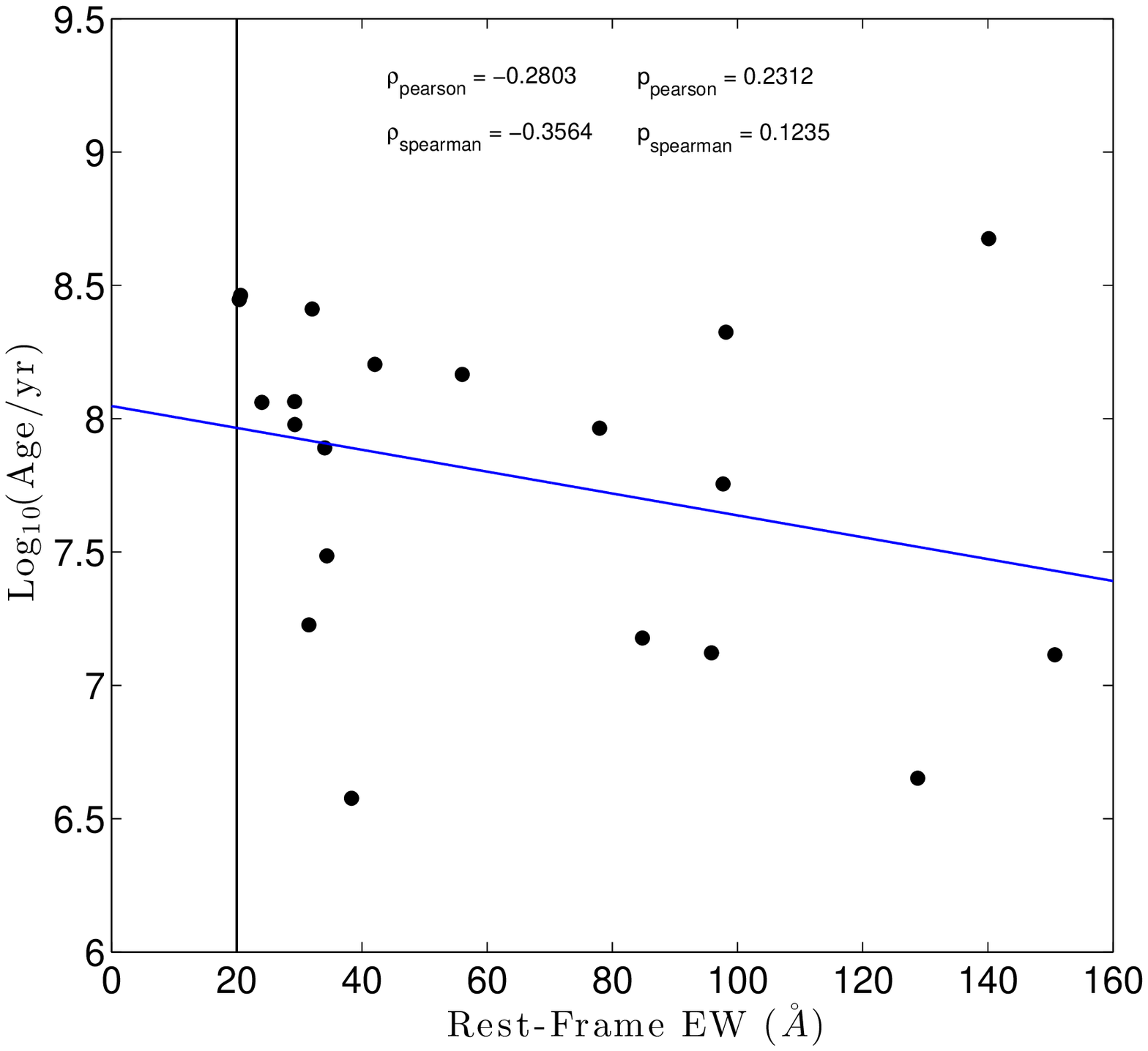}}
\subfloat[]{\includegraphics[width=2.2in]{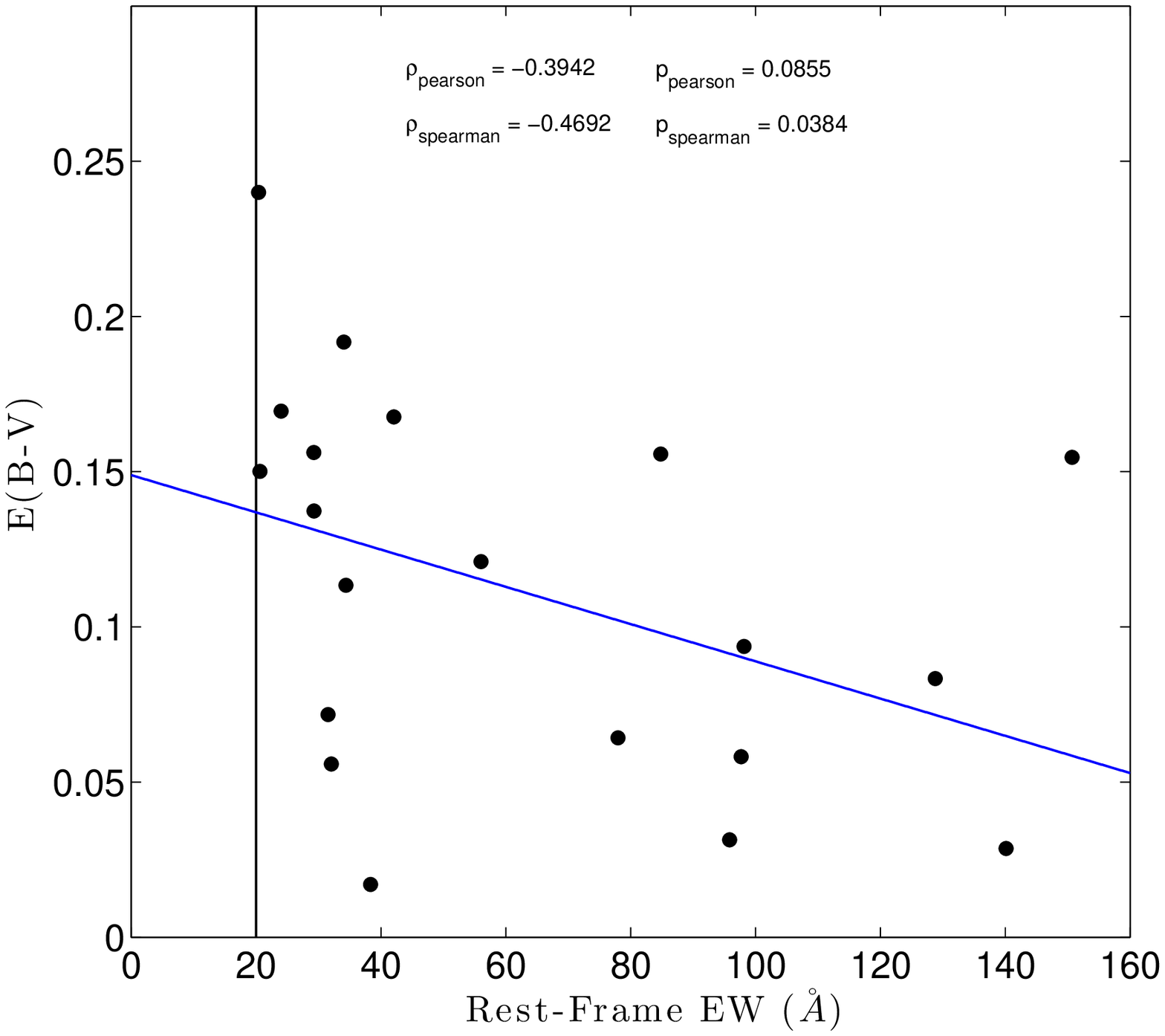}}
\caption{These plots show the relationships between (a) stellar mass
  and Ly$\alpha$ EW, (b) age and Ly$\alpha$ EW, and (c) E(B-V) and
  Ly$\alpha$ EW of the individual objects.\label{fig6}}
\end{figure*}

\subsection{Correlations with Equivalent Width}

The first panel of Figure \ref{fig6} shows stellar mass versus the rest-frame
Ly$\alpha$ equivalent width (EW).  There is a rough upper envelope
on the points such that no objects exhibit rest-frame $EW>60$
Angstroms and stellar mass$>10^9 M_{\odot}$.  Because EW was measured
via narrow-band $3727 \AA$ and broad-band U,B photometry
\citep{guaitaetal10}, there is no direct selection effect on stellar
mass.  Hence the lack of high-stellar-mass, high-EW objects implies
that galaxies with high stellar mass also have high rest-UV
luminosity.  This should be reflected via a related dearth of objects
with high stellar mass and low SFR i.e. low specific SFRs.

The second panel of Figure \ref{fig6} shows stellar population ages
versus rest-frame EW.  Here, one might expect to see a strong
correlation, as the Ly$\alpha$ EW of a starburst is a strongly
decreasing function of age, and for continuous star formation rate the
correlation remains strong \citep{shapleyetal03}. The observed lack of
correlation implies that Ly$\alpha$ radiative transfer is not trivial;
however, it could also be produced by a disconnect between our
measured ages and the true age of the starbursting population caused
by our assumption of a single stellar population with constant SFR.

The third panel of Figure \ref{fig6} shows dust reddening, E(B-V), versus
rest-frame EW.  Though statistically correlated, the correlation is
weak. One could expect that dustier galaxies would more effectively
quench Ly$\alpha$ photons to produce lower EWs. The lack of strong
correlation therefore implies that Ly$\alpha$ radiative transfer is
working to prevent Ly$\alpha$ photons from repeated resonant scatters
and hence prevents E(B-V) from strongly affecting the resulting
Ly$\alpha$ luminosity. This is a particularly interesting point due to
the fact that this effect is not seen in other recent LAE studies, but
is seen in some local group studies, such as \citet{giavaliscoetal96a}.

\begin{figure*}[ht]
\centering
\subfloat[]{\includegraphics[width=2.2in]{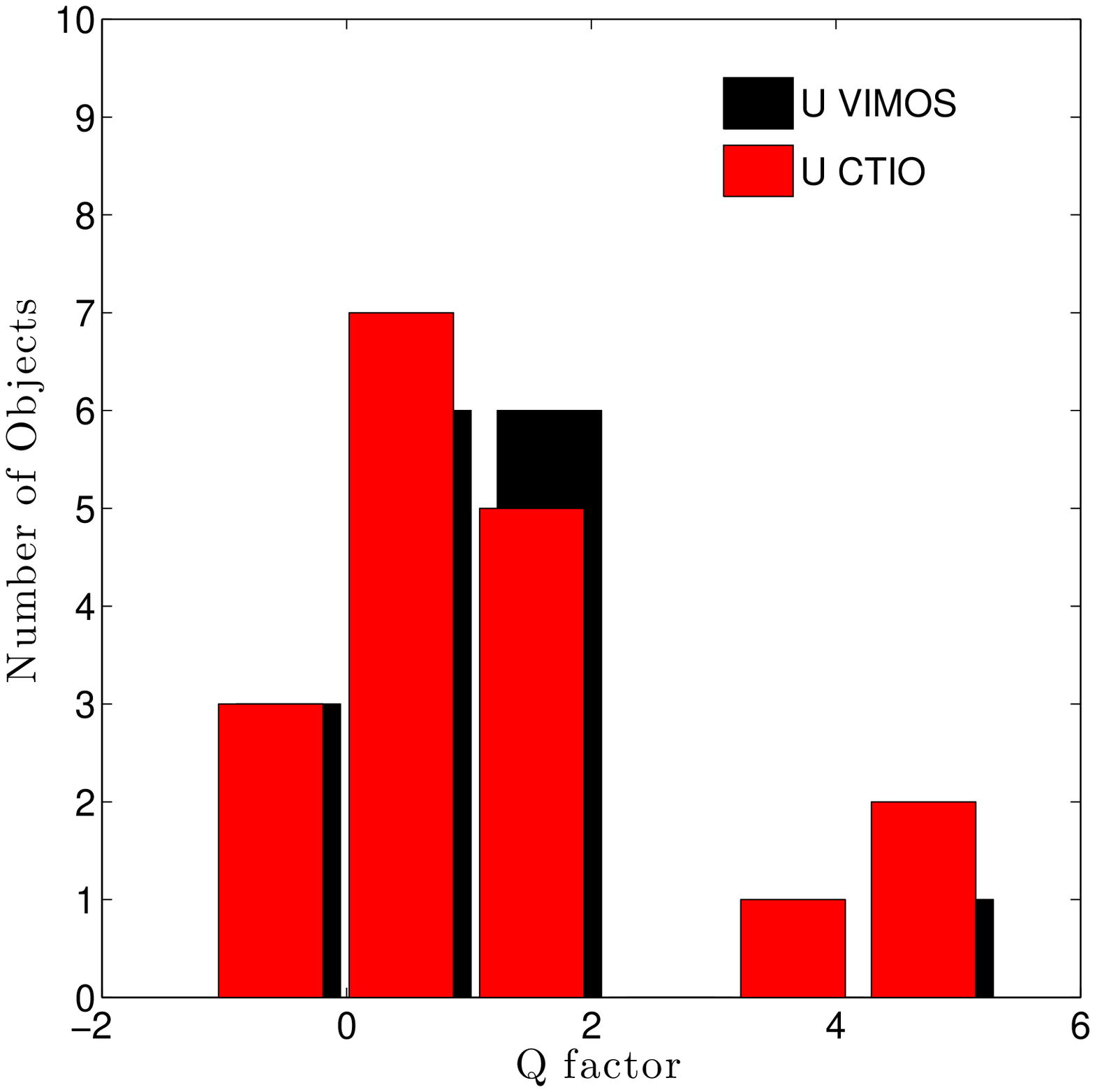}}
\subfloat[]{\includegraphics[width=2.2in]{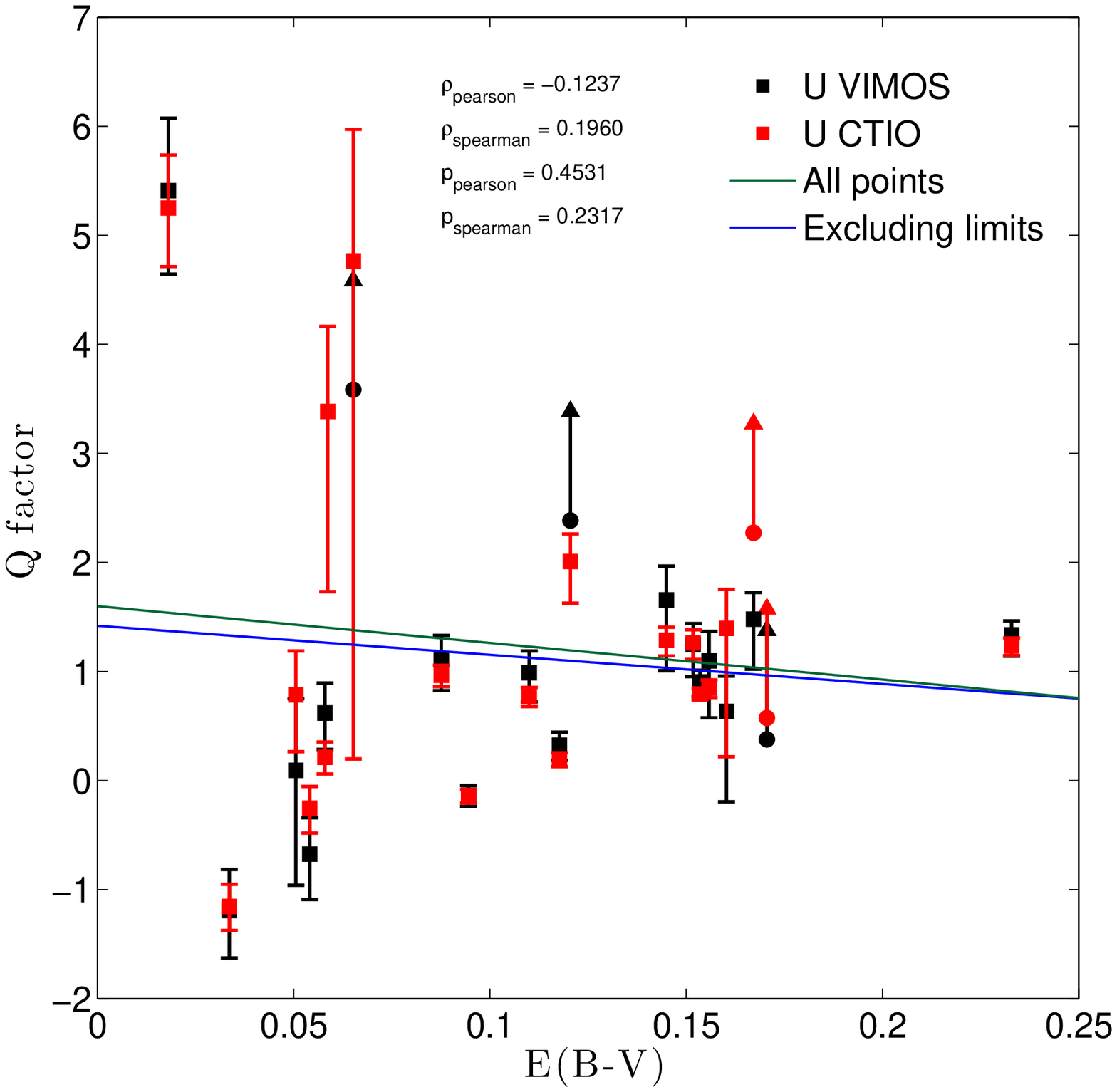}}
\subfloat[]{\includegraphics[width=2.2in]{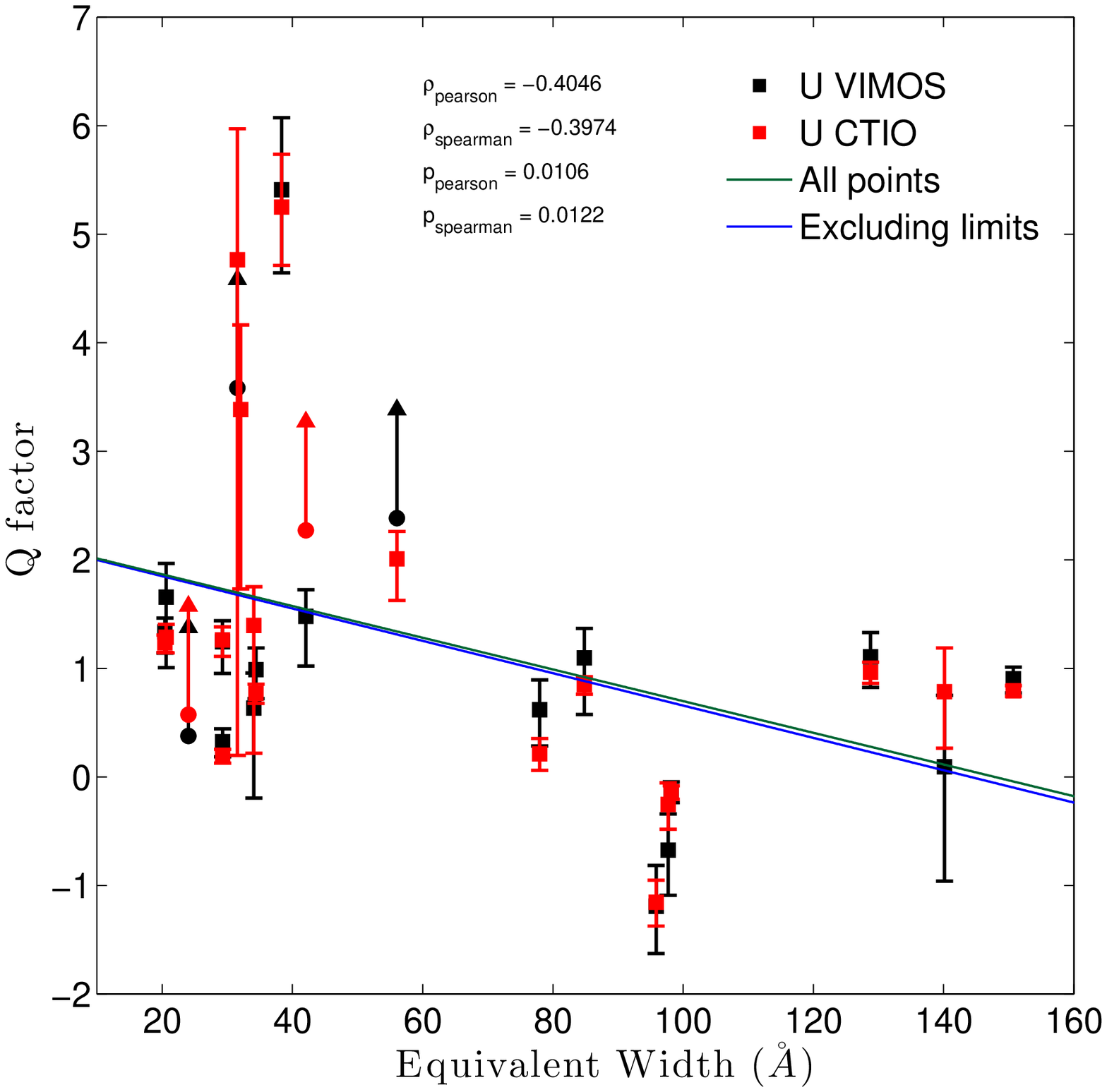}}
\caption{These plots show (a) the distribution of Ly$\alpha$ $q$
  factors for each of the individual objects, (b) the $q$ factor -
  E(B-V) relation, and (c) $q$ factor - Ly$\alpha$ EW relation for
  the individual objects.\label{fig7}}
\end{figure*}

\subsection{The Ly$\alpha$ q Factor}

As mentioned above, we excluded the U-band data from SED-fitting due
to this likelihood of Ly$\alpha$ emission making a significant
contribution to this broad-band photometry and the impossibility of
properly accounting for Ly$\alpha$ radiative transfer in our SED
templates.  However, this makes the U-band data ideal for an
estimation of Ly$\alpha$ $q$ factors, which represent the extent to
which Ly$\alpha$ emission from a galaxy has been enhanced ($q>1$) or
quenched ($q<1$) along the line-of-sight to Earth. As was first done
by \citet{finkelsteinetal08a}, we define $q=\tau_{Ly\alpha} /
\tau_{\lambda=1216}$, where $\tau_{\lambda} =k_{\lambda} E(B-V)/1.086$
and $k_{\lambda}$ follows a \citet{calzettietal00} dust attenuation
law. The $q$ factor has typically been measured from the same
narrow-band imaging used to select LAEs, leading to a selection
against low values of $q$. The deep U-band data in the CANDELS catalog
allows us to make a measurement of $q$ from broad-band imaging
sensitive to Ly$\alpha$ that is uncorrelated with the original
selection of the LAE sample.  A $q$ factor of unity implies that
Ly$\alpha$ photons see the same dust column as UV continuum photons,
whereas a $q$ factor of zero points to Ly$\alpha$ photons seeing no
dust extinction whatsoever. A negative $q$ factor would imply
Ly$\alpha$ enhancement by a top-heavy IMF, clumpy dust, or anisotropic
radiative transfer.

Panel (a) of Figure \ref{fig7} shows that most objects have $q$ factors between
zero and one, as expected. In a few of our objects, noise in the
U-band photometry results in very low, even negative, values of
$q$. There is no statistically significant correlation with E(B-V) in
panel (b), but panel (c) shows a correlation with EW in the expected
direction.

\subsection{Are LAEs on the SFR-$M_{*}$ ``Main Sequence''?}

\begin{figure}[t]
\centering
\includegraphics[scale=0.45]{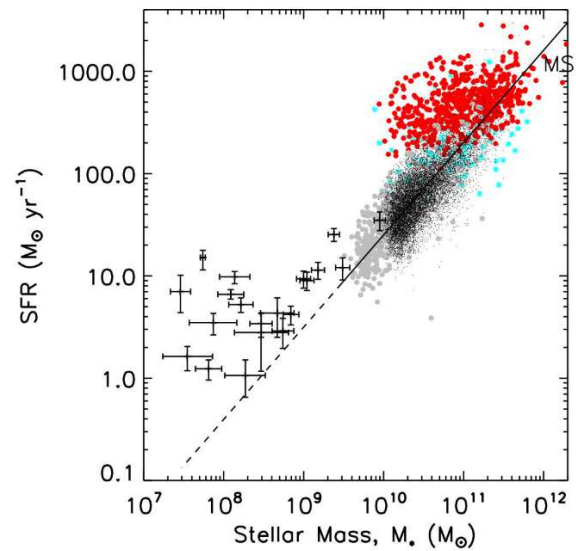}
\caption{LAEs at z = 2.1 compared to the galaxy main sequence.  Star
  Formation Rate (SFR) vs. stellar mass (M$_*$) for the LAE
  sample are shown as black points, with error bars.  For comparison,
  the SFR-M$_*$ relation at $1.5 < z < 2.5$ (galaxy main sequence, MS)
  is illustrated as the solid, black line (Daddi et al. 2007).
  Individual galaxy measurements reported in Rodighiero et al. (2011)
  are also shown: Herschel-PACS data for the COSMOS field (red
  circles) and GOODS-S (cyan squares), BzK-GOODS galaxies (black
  circles) and BzK-COSMOS galaxies (black points).\label{fig8}}
\end{figure}

Lastly, in Figure \ref{fig8}, we explore the behavior of the individual object
LAEs in the SFR - stellar mass relation. Our individual object LAEs
appear to lie systematically above the galaxy main sequence (MS),
implying larger star formation rates than expected for galaxies of
their mass. Interestingly, the scatter seems to increase toward faint
masses. This could happen because a star formation episode of a few
$M_{\odot} / yr$ will disturb a lower mass galaxy much more than a
higher mass galaxy. In the analysis done in \citet{ibaretal13} and
seen in their Figure 8, H$\alpha$ emitters at $z=1.47$ also fall above
the Main Sequence, though that study probes much higher masses. One
should note that all other studies shown on this plot used the same
(Salpeter) IMF, BC03 models, and constant star formation histories as
used in this study, ruling out a systematic offset in SFR or $M_{*}$
due to varying assumptions.

\subsection{Stacking Results}

In order to better discuss how well the individual object SED fitting
parameters agree with that of the various stacking techniques, three
rows were added to Table 1, containing the average, median, and
$1-\sigma$ scatter of the individual object parameters. For
simplicity, we define the failure of a stacking technique in a
particular parameter to be when the average (median) of the individual
results resides outside of the stacked parameter's error region. Using
this definition, the median \textit{HST} image stack fails in stellar
mass, the average \textit{NB} image stack fails in all three
parameters, and the median \textit{NB} image stack fails in stellar
mass. The average flux stack, median flux stack, and average
\textit{HST} image stack match the individual object results in all
three SED fitting parameters. It should be stressed that we do not
make direct comparisons between median stacks and average parameter
values (or vice versa). The spread of individual object and stacked
parameters can be seen in Figure \ref{fig5}. While it is not entirely
surprising that the NB-centered image stacks perform less well than
the \textit{HST}- centered image stacks, failure of any
\textit{HST}-centered stacking results is striking, since stacking
using the CANDELS established positions for the individual objects is
a best-case scenario for stacking accuracy. Lower precision position
estimates would provide a substantial amount of error in any stacking
analysis. It is possible that this discrepancy in the median image
stack could be due to a majority of objects receiving contaminating
flux from nearby objects in the IRAC infrared bands.

\begin{table*}[ht!]
  \center
  \caption{}
  \footnotesize
  \begin{tabular}{lcccc}
 \hline
\hline
CANDELS ID & M$^*$ ($10^9$ M/M$_{\odot}$) & Age (Gyr) & E(B-V) & best fit $\chi^2$/d.o.f \\
\hline
\hline 
205 (Clean) & $2.474^{+0.328}_{-0.446}$ & $0.115^{+0.032}_{-0.035}$ & $0.169^{+0.014}_{-0.012}$ & 81.9/8 \\ 
991 (Clean) & $0.407^{+0.353}_{-0.118}$ & $0.095^{+0.259}_{-0.041}$ & $0.137^{+0.018}_{-0.060}$ & 18.2/8 \\ 
7097 (Clean) & $0.971^{+0.250}_{-0.158}$ & $0.116^{+0.070}_{-0.032}$ & $0.156^{+0.013}_{-0.022}$ & 111.3/8 \\ 
9125 (Clean) & $0.280^{+0.052}_{-0.176}$ & $0.473^{+0.138}_{-0.394}$ & $0.029^{+0.059}_{-0.011}$ & 20.1/8 \\ 
10220 (Clean) & $0.717^{+0.154}_{-0.164}$ & $0.211^{+0.100}_{-0.083}$ & $0.094^{+0.020}_{-0.019}$ & 151.0/8 \\ 
11824 (Clean) & $0.023^{+0.050}_{-0.005}$ & $0.013^{+0.052}_{-0.004}$ & $0.031^{+0.022}_{-0.018}$ & 14.0/8 \\ 
12369 (Clean) & $0.064^{+0.030}_{-0.020}$ & $0.057^{+0.051}_{-0.024}$ & $0.058^{+0.013}_{-0.022}$ & 13.4/8 \\ 
14889 (Clean) & $0.101^{+0.078}_{-0.016}$ & $0.015^{+0.017}_{-0.003}$ & $0.156^{+0.011}_{-0.010}$ & 13.8/8 \\ 
17254 (Clean) & $8.518^{+2.021}_{-0.893}$ & $0.280^{+0.182}_{-0.063}$ & $0.240^{+0.016}_{-0.029}$ & 189.3/8 \\ 
22689 (Clean) & $0.286^{+0.118}_{-0.073}$ & $0.092^{+0.093}_{-0.036}$ & $0.064^{+0.015}_{-0.029}$ & 5.0/8 \\ 
24168 (Clean) & $1.116^{+0.229}_{-0.219}$ & $0.147^{+0.070}_{-0.051}$ & $0.121^{+0.019}_{-0.019}$ & 61.0/8 \\ 
2210 & $1.533^{+0.289}_{-0.281}$ & $0.160^{+0.069}_{-0.052}$ & $0.168^{+0.018}_{-0.019}$ & 28.0/8 \\ 
3826 & $0.024^{+0.014}_{-0.003}$ & $0.004^{+0.004}_{-0.002}$ & $0.083^{+0.034}_{-0.026}$ & 4.2/8 \\ 
6696 & $0.231^{+0.415}_{-0.095}$ & $0.078^{+0.581}_{-0.042}$ & $0.192^{+0.025}_{-0.094}$ & 11.3/8 \\ 
7408 & $0.053^{+0.006}_{-0.002}$ & $0.004^{+0.001}_{-0.001}$ & $0.017^{+0.011}_{-0.010}$ & 40.4/8 \\ 
11292 & $0.154^{+0.080}_{-0.038}$ & $0.031^{+0.027}_{-0.010}$ & $0.113^{+0.007}_{-0.014}$ & 14.1/8 \\ 
18836 & $0.124^{+0.089}_{-0.035}$ & $0.013^{+0.013}_{-0.004}$ & $0.155^{+0.017}_{-0.019}$ & 24.0/8 \\ 
21026 & $0.060^{+0.086}_{-0.023}$ & $0.017^{+0.041}_{-0.007}$ & $0.072^{+0.016}_{-0.029}$ & 4.9/8 \\ 
21553 & $0.572^{+0.184}_{-0.169}$ & $0.258^{+0.203}_{-0.134}$ & $0.056^{+0.035}_{-0.031}$ & 6.1/8 \\ 
25902 & $2.966^{+0.807}_{-0.454}$ & $0.290^{+0.212}_{-0.090}$ & $0.150^{+0.020}_{-0.030}$ & 94.0/8 \\
\hline
Average of Individual Results & 1.03 & 0.123 & 0.113 & \\
Median of Individual Results & 0.283 & 0.094 & 0.117 & \\
Scatter of Individual Results & 1.896 & 0.121 & 0.059\\    
\hline
Average Flux Stack & $1.050^{+0.796}_{-0.487}$ & $0.184^{+0.471}_{-0.126}$ & $0.146^{+0.042}_{-0.066}$ & 3.6/8 \\ 
Median Flux Stack & $0.511^{+0.347}_{-0.237}$ & $0.148^{+0.335}_{-0.104}$ & $0.108^{+0.048}_{-0.062}$ & 1.7/8 \\ 
Average HST Image Stack & $1.192^{+1.262}_{-0.448}$ & $0.309^{+0.873}_{-0.214}$ & $0.124^{+0.066}_{-0.059}$ & 22.5/8 \\ 
Median HST Image Stack & $0.810^{+0.451}_{-0.428}$ & $0.305^{+0.509}_{-0.245}$ & $0.108^{+0.075}_{-0.053}$ & 11.8/8 \\ 
Average NB Image Stack & $0.222^{+0.180}_{-0.106}$ & $0.018^{+0.044}_{-0.010}$ & $0.165^{+0.006}_{-0.057}$ & 43.5/8 \\ 
Median NB Image Stack & $0.828^{+0.396}_{-0.489}$ & $0.462^{+0.559}_{-0.401}$ & $0.101^{+0.103}_{-0.042}$ & 16.9/8 \\ 
\hline
Scaled Average Flux Stack & $0.525^{+0.280}_{-0.043}$ & $0.084^{+0.093}_{-0.017}$ & $0.105^{+0.025}_{-0.014}$ & 48.2/8 \\ 
Scaled Median Flux Stack & $0.382^{+0.165}_{-0.074}$ & $0.091^{+0.096}_{-0.028}$ & $0.131^{+0.013}_{-0.030}$ & 3.31/8 \\ 
\hline
Clean Average Flux Stack & $1.314^{+1.124}_{-0.753}$ & $0.206^{+0.635}_{-0.163}$ & $0.160^{+0.057}_{-0.076}$ & 4.6/8 \\ 
Clean Median Flux Stack & $0.639^{+0.182}_{-0.526}$ & $0.348^{+0.098}_{-0.331}$ & $0.048^{+0.104}_{-0.016}$ & 3.7/8 \\ 
Clean Average HST Image Stack & $0.088^{+0.140}_{-0.031}$ & $0.008^{+0.035}_{-0.004}$ & $0.140^{+0.028}_{-0.069}$ & 20.9/8 \\ 
Clean Median HST Image Stack & $0.049^{+0.309}_{-0.009}$ & $0.009^{+0.177}_{-0.004}$ & $0.089^{+0.062}_{-0.058}$ & 23.2/8 \\ 
Clean Average NB Image Stack & $0.095^{+0.126}_{-0.035}$ & $0.007^{+0.029}_{-0.004}$ & $0.150^{+0.023}_{-0.072}$ & 15.8/8 \\ 
Clean Median NB Image Stack & $0.079^{+0.214}_{-0.049}$ & $0.016^{+0.170}_{-0.012}$ & $0.131^{+0.042}_{-0.098}$ & 10.4/8 \\ 
\hline
\end{tabular}
\end{table*}

The average \textit{NB} image stack is poor at estimating the average
parameters for the sample, but other stacking techniques are
relatively reasonable. While the error bars for the stacks shown in
Figure \ref{fig3} include sample variance, the large dispersion of
properties is still not captured by any stacking method. As can be
seen in Table 1 and Figure \ref{fig5}, there is a large dispersion in
the values of mass, age and E(B-V) measured from individual objects,
and this dispersion is far larger than the measurement uncertainties.
As typically implemented, the stacking approach misses this
dispersion, and this is a significant weakness versus fitting
individual object SEDs.  In circumstances where stacking must be used,
however, there are statistical methods that can be used to infer the
scatter in properties within the population.  An upper limit to the
uncertainty in the parameters inferred for the stacked population can
be obtained by assuming that the dispersion in properties of the
underlying population is much larger than the photometric errors
(although one should then interpret a median stack carefully because
there may not exist a "typical object"). In this case, the scatter in
the parameters for a population of $N$ objects could be quantified as
$\sqrt{N}$ times the uncertainty found for their average-stacked SED;
this effectively turns the standard deviation of the average back into
the standard deviation of the population.  We do find that the sample
variance found during our bootstrap resampling dominates the flux
uncertainties on the stacked SED in most wavebands. However, applying
this method to the average flux stack values in Figure \ref{fig1}
would significantly overestimate the observed scatter in individual
object properties.  A more sophisticated approach would be to use
jacknife techniques \citep{lupton93} where stacks are made from each
set of $N-1$ galaxies and the resulting variance in best-fit
parameters is turned into an estimate of the scatter of the
population.

\subsection{Scaled Flux Stacks}

While straightforward calculations of average (median) stacks have
been standard in the literature, it is also possible to scale all
input fluxes (or images) to a common brightness e.g., in H-band,
before stacking.  This eliminates the influence of variation in
overall brightness and focuses the stacking on determining an average
(median) SED shape.  To test the efficacy of this approach, we
introduce {\em scaled average} and {\em scaled median} flux stacks.
The {\em scaled average (median)} flux stack was created by
determining the average (median) H-band flux and then multiplying each
individual input SED and its uncertainties by the factor needed to
match that value.
As with other stacks, we include uncertainties due to sample variance
by following the identical procedure in bootstrap simulations.  By
their definition, these new {\em scaled} average (median) stacked SEDs have
the same average (median) as the corresponding flux-stacked SEDs.
Their shapes turn out to be similar.  Nonetheless, there is
significantly less variance among the bootstrap samples due to having
removed the impact of brightness variations, and the {\em scaled} flux
stacks have significantly smaller uncertainties as a result.
In Figure \ref{fig5}, we see that these scaled flux stacks perform
quite well in tracing the properties of the sample, at least as well
as the simple flux-stacked SEDs but with significantly reduced
parameter uncertainties due to the reduced uncertainties in the SED.
We conclude that scaling is a superior method for producing stacked
SEDs and recommend its use in future studies.

\subsection{Clean Stacks}

As briefly discussed above, in an attempt to test whether or not IRAC
contamination could be a strong source of error, we compiled an
additional set of stacks using only the ten individual objects whose
CANDELS images were free of nearby sources of flux. This was
determined by a visual inspection of the photometric images of each
object. If an object's image contained nearby extraneous sources it
was not included in these ``clean'' stacks. Our parallel analysis with
these clean stacks produced results similar to the original
stacks. So, there is no evidence that IRAC contamination can account
for the discrepancies of our results. However, the clean stack
analysis lowers the sample size of an already sparse set of objects,
so our clean stacking analysis might suffer from significant
information loss. Clean stack results are included in Table 1.

The amount gained in a clean versus crowded analysis is
questionable. Figure \ref{fig2} shows the average and median stacked
images for the entire sample and the clean sample. Inspection of the
rightmost panel shows that a clean sample of 10 objects produces a
less reliable median. Rather than excluding crowded regions, the
CANDELS catalog photometry used to create our flux stacks utilizes
TFIT \citep{laidleretal07}, which seems to perform better than careful
qualitative crowdedness assessments by eye. This could cause the flux
stacks to be more accurate than image stacks.

\section{Discussion \& Conclusions}

Stacking techniques have been employed for studying low signal to
noise objects. However, some stacking methods may be more accurate at
modeling the parameters of a sample than others. Through fitting
individual SEDs of 20 LAEs at redshift 2.1 individually, and then
fitting 6 types of stacked SEDs comprised of the same 20 objects, we
have tested the validity of these methods. We find that median flux
stacking and average \textit{HST}-centered image stacking techniques
correctly represent the age and stellar mass of a sample, while the
median \textit{HST}-centered image stack performs less
adequately. This is a stark result, since our analysis serves as a
best-case scenario for stacking. Stacking objects with CANDELS defined
coordinates maximizes the accuracy of our stacking sample. We further
investigated this result by establishing a set of clean stacks, which
are comprised of objects free of nearby sources of flux, and with
simulations. The clean stacking analysis sheds little light on the
underlying reason for the discrepancies in image-stacked parameter
estimates of a sample of individual objects. Future stacking analyses
should be wary of disparities in analyses due to errors possibly
brought on by the stacking method itself. The simulations show that
median image stack photometry is biased considerably low when
centroiding errors are significant compared to the PSF. Though this
effect is not, in practice, catastrophic for either the \textit{NB} or
\textit{HST}- centered image stacks, it provides a note of caution. We
caution the reader of inconsistencies any stacking analysis may bring
forth, and recommend the use of a \textit{scaled} stacking method
where stacking is absolutely necessary.

Our main conclusions are the following: 

\begin{itemize}
   \item A lack of correlation between Ly$\alpha$ EW and age implies 
     either complicated radiative transfer mechanisms, or an
     inappropriate assumption of a constant SFR in a starbursting sample.
   \item The narrow distribution of $q$ values peaking near one and
     lack of correlation between $q$, and E(B-V) implies
     that radiative transfer mechanisms seem to be working to prevent
     Ly$\alpha$ photons from resonantly scattering in dusty regions.
   \item Our sample of LAEs lies systematically above the SFR-stellar
     mass relation galaxy ``main sequence'' and shows an increase in
     scatter above this relation at low mass. This may be caused by
     ongoing starbursts in these galaxies causing a greater excursion
     in their specific star formation rates.
   \item Though some types of stacking represent the average and
     median properties of a sample well, the large dispersion of
     individual object properties is obscured by stacking. We
     recommend a new approach using an H-band scaled median or average
     flux stack, which reduces uncertainties significantly.
\end{itemize}

\section*{Acknowledgements}
We would like to thank both the CANDELS and MUSYC collaborations for
making this work possible. This work is based on observations taken by
the CANDELS Multi-Cycle Treasury Program with the NASA/ESA HST, which
is operated by the Association of Universities for Research in
Astronomy, Inc., under NASA contract NAS5-26555. These observations are
associated with programs GO-9352, GO-9425, GO-9583, GO-9728, GO-10189,
GO-10339, GO-10340, GO-11359, GO-12060, and GO-12061. Observations
have been carried out using the Very Large Telescope at the ESO
Paranal Observatory under Program ID(s): LP168.A-0485. This work is
also based in part on observations made with the Spitzer Space
Telescope, which is operated by the Jet Propulsion Laboratory,
California Institute of Technology, under a contract with
NASA. Support for Program numbers 12060.57 \& 12445.56 was provided by
NASA through a grant from the Space Telescope Science Institute, which
is operated by the Association of Universities for Research in
Astronomy, Incorporated, under NASA contract NAS5-26555. The Institute
for Gravitation and the Cosmos is supported by the Eberly College of
Science and the Office of the Senior Vice President for Research at
the Pennsylvania State University. This material is based on work
supported by the National Science Foundation under CAREER grant
AST-1055919 awarded to Eric Gawiser. This research has made use of
NASA’s Astrophysics Data System.

\bibliographystyle{apj}
\bibliography{refs}

\end{document}